\documentclass[pra,aps,superscriptaddress,balancelastpage,onecolumn]{revtex4}
\usepackage{bbding}
\usepackage[utf8]{inputenc}
\usepackage{amsthm}
\usepackage{dcolumn}
\usepackage{bm}
\usepackage{subfigure}
\usepackage{epsfig,graphicx,times}
\usepackage{amstext}
\usepackage{amsmath}
\usepackage{amssymb}
\usepackage{graphicx}
\usepackage{latexsym}
\usepackage{bm}
\usepackage{graphicx,epstopdf}
\usepackage[dvipsnames]{xcolor}
\usepackage{epsfig}
\usepackage{graphics}
\usepackage{tabularx}
\usepackage{subfigure}
\usepackage[mathlines]{lineno}
\usepackage{color}
\providecommand{\U}[1]{\protect\rule{.1in}{.1in}}
\providecommand{\U}[1]{\protect\rule{.1in}{.1in}}

\begin{document}

\title{Enhancement of Entanglement via Josephson Parametric Amplifier in a Dual Cavity-Magnon System}
\author{Abdelkader Hidki}
\affiliation{LPTHE, Department of Physics, Faculty of Sciences, Ibn Zohr University, Agadir, Morocco}
\author{Abderrahim Lakhfif}
\affiliation{LPHE-Modeling and Simulations, Faculty of Science, Mohammed V University in Rabat, Rabat, Morocco}
\author{Mostafa Nassik}
\affiliation{LPTHE, Department of Physics, Faculty of Sciences, Ibn Zohr University, Agadir, Morocco}
\author{Rizwan Ahmed}
\affiliation{Physics Division, Pakistan Institute of Nuclear Science and Technology
(PINSTECH), P. O. Nilore, Islamabad 45650, Pakistan.}
\author{Amjad Sohail}
\email{amjadsohail@gcuf.edu.pk}
\affiliation{Department of Physics, Government College University, Allama Iqbal Road, Faisalabad 38000, Pakistan}
\begin{abstract}
In the two microwave (MW) cross-shaped cavity magnon system, we describe a method to produce multipartite entanglement and quantum steering. To achieve squeezed states of the magnons, a Josephson parametric amplifier (JPA) creates a squeezed vacuum field that drives the two cavities. We theoretically demonstrate that the cavity-cavity entanglement can be generated at the resonance point, however, increasing the cavity and magnon decay rates generate the cavity-magnon entanglement. By changing the squeezing parameter and increasing the decay rates, we can transfer the cavity-cavity entanglement to cavity-magnon entanglement. Furthermore, the cavity-cavity entanglement (survive up to 2.8K) not only found to be much stronger but also more robust as compared to cavity-magnon entanglement (survive up to 0.4K). More importantly, the genuine photon-magnon-photon tripartite entanglement  could be achieved, which is robust against the thermal fluctuations and depends strongly on squeezing parameter. Furthermore, for current dual cavity-magnon system, two-way quantum steering is found when the optomagnonical couplings are equal. The current study offers a straightforward and practical method for achieving multipartite quantum correlations.
\end{abstract}

\maketitle

\section{Introduction}
The investigation of macroscopic quantum states has received increasing attention due to significant advances in experimental technology, especially since the proposal of Schr\"{o}dinger's cat states\cite{1}. Notably, cavity optomechanics (COM) \cite{2,3}, which highlights the complex interaction between electromagnetic fields and mechanical motion through radiation pressure, and thereby presents a possibility of creating macroscopic quantum states. Over the past decade, considerable advancement have been carried out in optomechanical systems to generate macroscopic quantum states in massive mechanical oscillators \cite{4,5}. These achievements incorporate the successful realization of finding entangled states involving both mechanical oscillators and electromagnetic fields \cite{6}, as well as entangled states between two mechanical oscillators \cite{7,8,9,10}. Moreover, researchers have successfully produced  superposition states \cite{11,12} and squeezed states \cite{13} for mechanical motion, etc. In addition, COM has enabled the generation of nonclassical states, such as Fock states \cite{14}, superposition states \cite{15},  cat states \cite{16} and entangled states \cite{17,18}, in macroscopic mechanical resonators by controlled coupling with superconducting qubits.

Recently, the cavity-magnonics system, with the ferrimagnetic materials such as the famous yttrium-iron-garnet (YIG) sphere \cite{19}, has emerged as a prominent area of research \cite{20}. In these quantum systems, which are based on cavity quantum electrodynamics (QED), basic excitations brought on by uniform magnetic fields lead to the formation of magnons (Kittel mode) in a YIG crystal \cite{21}. These magnons can be directly coupled to MW photons \cite{22,23,24}, optical photons \cite{25,26}, and mechanical phonons \cite{27} via the processes of magnetic-dipole interaction, magneto-optical effects, and magnetostrictive interaction, respectively. This compelling area of research has uncovered a multitude of intriguing phenomena, including the exceptional points \cite{28}, Kerr effect \cite{29}, non-reciprocity transmission \cite{30}, magnon-induced absorption \cite{31},  magnon-induced transparency \cite{32}, coherent MW-to-optical conversion \cite{33}, magnon squeezing by two-tone driving of a qubit \cite{34}, and quantum entanglement of magnons \cite{35,36,37,38}.
	
Along with the coupling between MW and magnons, the latter can also interact with the deformation vibration phonons of the YIG sphere through nonlinear magnetostrictive interaction, giving rise to what is called cavity magnomechanics (CMM) \cite{39}. 	This interaction opens up various fascinating possibilities, such as photon-magnon-phonon entanglement, steering and magnon-squeezing states \cite{40,41,42,43,44,45,46,47,48,49,50}, generation of magnon laser or chaos \cite{51,52}, ultra-slow light engineering \cite{32,53}, use of coherent feedback loop \cite{54,55}, realization of ground-state cooling of mechanical vibrations mode \cite{56,57}, magnon-assisted photon-phonon conversion \cite{58} and  quantum state retrieval and storage \cite{59}, among several other applications. An important observation is
is how frequently squeezed-reservoir engineering has been used to examine entanglement and squeezing in different optomechanical and magnomechanical systems \cite{60}. This involves utilizing broadband squeezed light as a squeezed reservoir to drive the cavity, and it has found extensive application across different systems \cite{10,61,62}. Particularly, single-mode squeezed light is employed to generate Einstein-Podolsky-Rosen (EPR) quantum steering \cite{63}, and magnon-magnon correlation \cite{35} in the cavity-magnon system. Furthermore, two-mode squeezed microwave fields are also used to observe a strong correlation between two magnon modes \cite{36}, the vibrational modes of two YIG spheres \cite{44}, and manipulate one-way Gaussian steering \cite{64}.

In this manuscript, we propose an efficient scheme to achieve bi-(tri)partite entanglement and quantum steering among a magnon and two MW fields. Several suggestions have been made previously, proposing distinct mechanisms for preparing entangled states involving two MW cavities, including the nonlinearity of magnetostrictive force \cite{43,45,46}, the Kerr nonlinearity \cite{65}, and the  optical parametric amplifier \cite{66}. Here, we propose a different approach where squeezed light is used to enhance genuine bi-(tri-)partite entanglement between a magnon and two MW fields in a cavity-magnon system.

It is worth noting that the first experimental demonstration of spatially distant two-mode squeezed states of MW light was achieved in Ref. \cite{67} by E. Flurin et. al.. Using a flux-driven Josephson parametric amplifier (JPA) to prompt the squeezed vacuum MW fields \cite{68,69}, the two MW cavity photons and the magnon can be entangled simultaneously.
In addition, a variety of factors employed in the current system can be changed to affect multipartite entanglement and quantum steering. Moreover, a true photon-magnon-photon tripartite entanglement is present in the current dual cavity-magnon system, and it can be realized with experimentally realizable parameters.
We show that for the current cavity-magnon system, the squeezing parameter is
cruicial for manipulating and controlling both multipartite entanglement and quantum steering within the system. Additionally, the generated entanglement exhibit more robustness against changes in the environmental temperature.

The remainder of the article is structured as follows. The model Hamiltonian of the current cavity-magnon system is described in Section 2. We use the standard Langevin approach in Section 3 to examine the system's dynamics and obtain a drift matrix. Additionally, in Section 4, we analytically examine bipartite, tripartite, and quantum steering while we present conclusion in Sect. 5.
\section{The Model}
As shown in Fig. 1, the cavity-magnon system comprise a magnon $m$ and the two MW cavity modes, $c_{1}$ and $c_{2}$.
The two cavities are arranged in a cross-shaped and simultaneously pumped by a two-mode squeezed vacuum MW field induced by JPA \cite{69}.
The collective excitations of many spins inside a YIG sphere serve as the physical representation of a quasiparticle known as a magnon \cite{41}.
The magnetic dipole interactions strengthen the optomagnonic coupling strength between the two MW cavity modes and the magnon because of the high spin density in the YIG sphere \cite{22,23,70}. The YIG sphere is positioned inside the cross-shaped cavities where there exist maximum magnetic fields of the two MW cavity modes. The magnetic field of the cavities $c_{1}$ ($c_{2}$) is along x-(y-)directions. Furthermore, a bias magnetic field is also applied which is perpendicular to both the magnetic field of the cavity modes. In addition, we also consider a very small sized YIG sphere as compared to wavelengths of the MW cavities, therefore, we can safely neglect the effects of radiation pressure induced by MW fields \cite{41}. The Hamiltonian of the current cavity-magnon system reads as:
\begin{figure}[tbh]
	\centering
	\includegraphics[width=0.78\columnwidth,height=3.2in]{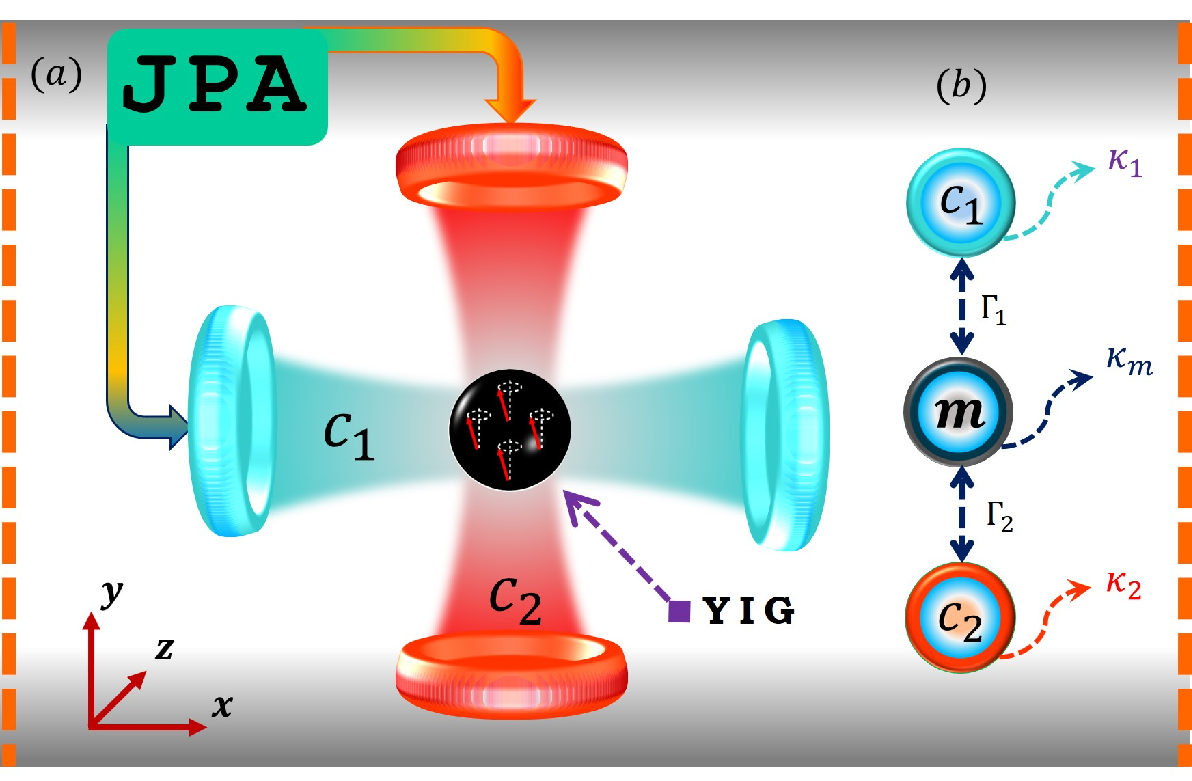} \centering
\caption{(\textbf{a}) Graphical representation of the proposed system: A magnon in a YIG sphere is positioned inside two cross-shaped MW cavities near the maximum magnetic fields of the cavity modes and simultaneously in a uniform bias magnetic field. The magnetic dipole interaction is responsible to couple two MW fields $c_1$ and $c_2$ with a magnon mode $m$ in the YIG sphere. JPA induces a two-mode squeezed vacuum to simultaneously drive the two cross-shaped cavities (\textbf{b}). The $k$th MW cavity mode, with dissipation rate $\kappa_k$ ($k=1,2$), couples with the magnon mode $m$, with decay rate $\kappa_m$, by the coupling strength $\Gamma_k$. }
\end{figure}

\begin{eqnarray}
	H &=&\hbar\sum_{k=1}^{2}\omega _{k}c_{k}^{\dag }c_{k}+\hbar\omega _{m}m^{\dag }m+\hbar\sum_{k=1}^{2}\Gamma_{k}\left( c_{k}m^{\dag
	}+c_{k}^{\dag }m\right),
\end{eqnarray}%
where $c_{k}$ and $\left( c_{k}^{\dag }\right) $ are annihilation and creation of the $k$th cavity mode while $m$ and $\left( m^{\dag }\right)$ represent the annihilation creation operator for the magnon mode. In addition, both cavity and magnon modes satisfy the following relation, $[z,z^{\dag}]=1$ ($z=m, c_k$). In addition, $\omega _{k}$ ($\omega _{m}$) denotes the resonance frequency of the $k$th cavity (magnon) mode. $\omega _{m}$ can be tuned via $\omega _{m}=\gamma_{0}B$, where $B$ is the external bias magnetic field and $\gamma _{0}$ denoted the gyromagnetic ratio. In addition, $\Gamma_{k}$ defines the optomagnonical coupling between the magnon mod and the $k$th cavity. Furthermore, $\Gamma_{k}$ is taken much larger than the cavity decay rate $\kappa_k$ and magnon decay rates $\kappa_m$, i.e., $\Gamma_{k}>\kappa_{k}$, $\kappa_{m}$, leading to strong coupling regime \cite{22,23,44,70}.

In the rotating frame at the drive frequency $\omega _{l}$ (the frequency of the output of the two-mode squeezed field generated by the JPA), the
Hamiltonian of the current cavity-magnon system takes the form:
\begin{eqnarray}
	H/\hbar &=&\sum_{k=1}^{2}\Delta _{k}c_{k}^{\dag }c_{k}+\Delta _{m}m^{\dag }m+\sum_{k=1}^{2}\Gamma_{k}\left( c_{k}m^{\dag }
	+c_{k}^{\dag }m\right),
\end{eqnarray}
where  $\Delta _{k}$ and $\Delta_{m}$ are the cavity and magnon detunings, and are  $\Delta _{k}=\omega _{k}-\omega _{l}$ ($k=1,2$) and $\Delta
_{m}=\omega _{m}-\omega _{l}$.

\section{Dynamics of the system}\label{sec3}
We utilize the standard quantum Langevin formalism to describe the system's dynamics. By incorporating the dissipation's of the cavity-magnon system, we arrive to the following set of quantum Langevin equations (QLEs) in the rotating frame:
\begin{eqnarray}
	\dot{c}_{k} &=&-\left(\kappa_{k}+ i\Delta _{k}\right) c_{k}-i\Gamma_{k}m+\sqrt{2\kappa_{k}}c_{k}^{in},\;\; k=1,2  \label{LED}\\
	\dot{m} &=&-\left(\kappa_{m}+ i\Delta _{m}\right)
	m-i\sum_{k=1}^{2}\Gamma_{k}c_{k} +\sqrt{2\kappa_{m}}m^{in},  \label{LEE}
\end{eqnarray}%
where $\kappa_{k}$ and $\kappa_{m}$ are the dissipation rates of the $k$th cavity mode and magnon mode, respectively. Additionally, the zero mean input noise operators for the MW cavities and the magnon modes, respectively, are $c_{k}^{in}$ and $m^{in}$. These noise operators incorporate the squeezed field's driving of the two cavities and meet the following time-domain correlation functions \cite{71}:
\begin{eqnarray}\label{corrF1}
	\left\langle c_{k}^{in\dag}(t)c_{k}^{in}(t^{\prime })\right\rangle &=&N\delta(t-t^{\prime }),\quad k=1,2  \\
	\left\langle c_{k}^{in}(t)c_{k}^{in\dag }(t^{\prime })\right\rangle &=&(N+1)\delta (t-t^{\prime }),  \\
	\left\langle c_{k}^{in}(t)c_{l}^{in}(t^{\prime })\right\rangle &=&\left\langle c_{k}^{in\dag}(t)c_{l}^{in\dag}(t^{\prime })\right\rangle=M\delta (t-t^{\prime }),\quad k\neq l=1,2 \\
	\left\langle m^{in\dag }(t)m^{in}(t^{\prime })\right\rangle &=&n_{m}\delta (t-t^{\prime }), \ \ \ \  \left\langle m^{in}(t)m^{in\dag }(t^{\prime})\right\rangle =(n_{m}+1)\delta (t-t^{\prime }),  \label{corrF2}
\end{eqnarray}
where $M={\rm sinh} (r){\rm cosh} (r)$ and $N={\rm sinh}^2 (r)$, $r$ being the squeezing parameter of the two-mode squeezed vacuum field, generated by JPA. $n_{m}=[\exp (\frac{\hbar\omega_{m}}{k_{B}T})-1]^{-1}$ is the equilibrium mean thermal magnon number, with $k_{B}$ is the Boltzmann constant and $T$ the environmental temperature \cite{40}.

To investigate the quantum correlations in the cavity-magnon system, we first derive the quantum fluctuations of the magnon and MW cavity modes. We then decompose each operator as: $z=\left\langle z\right\rangle +\delta z$ $(z=c_{k},m) $, where $\left\langle z\right\rangle$ ($\delta z$) is the average value (quantum fluctuation). In this way, we obtain the linearized set of QLEs explaining the dynamics of the quantum fluctuations:
\begin{eqnarray}
	\delta\dot{c}_{k} &=&-\left(\kappa_{k}+ i\Delta _{k}\right) \delta c_{k}-i\Gamma_{k}\delta m+\sqrt{2\kappa_{k}} c_{k}^{in},  \label{LLED}\\
	\delta\dot{m} &=&-\left(\kappa_{m}+ i\Delta _{m}\right)\delta m-i\sum_{k=1}^{2}\Gamma_{k}\delta c_{k} +\sqrt{2\kappa_{m}} m^{in}.  \label{LLEE}
\end{eqnarray}
In addition, the average value of each dynamical operator is given by:
\begin{eqnarray}
	\left\langle c_{k}\right\rangle &=&\frac{-i\Gamma _{k}\left\langle
		m\right\rangle }{\left(\kappa _{k}+ i\Delta _{k}\right) },   \\
	\left\langle m\right\rangle &=&\frac{-i\sum_{k=1}^{2}\Gamma
		_{k}\left\langle c_{k}\right\rangle}{\left(\kappa _{m}+ i\Delta _{m}\right) }
\end{eqnarray}

We now switch to the quadrature form of the QLEs by introducing the quadratures of the quantum fluctuations defined as $\delta m=\frac{1}{\sqrt{2}}(\delta x+i\delta y)$ and $\delta c_{k}==\frac{1}{\sqrt{2}}(\delta X_{k}+i\delta Y_{k})$
Therefore, Eqs. (\ref{LLED}) and (\ref{LLEE}) can be written in the following compact matrix form:
\begin{equation}
	\dot{\mathcal{B}}(t)=\mathcal{M}\mathcal{B}(t)+\mathcal{N}_{{noise}}(t),
\end{equation}%
where quantum fluctuation $\mathcal{B}(t)$ and input noise $\mathcal{N}_{{noise}}(t)$ vectors are given by:
\begin{eqnarray}
	\mathcal{B}(t)&=&[\delta x(t),\delta
	y(t),\delta X_{1}(t),\delta Y_{1}(t),\delta X_{2}(t),\delta Y_{2}(t)]^{\rm T}, \\
	\mathcal{N}_{{noise}}(t)&=&[\sqrt{2\kappa_{m}}x^{in}(t), \sqrt{2\kappa_{m}}y^{in}(t),\sqrt{2\kappa_{1}}X_{1}^{in}(t),\sqrt{2\kappa_{1}}Y_{1}^{in}(t),%
	\sqrt{2\kappa_{2}}X_{2}^{in}(t), \sqrt{2\kappa_{2}}Y_{2}^{in}(t)]^{\rm T}.
\end{eqnarray}
Furthermore, the drift matrix $\mathcal{M}$ is given by:
\begin{equation}
	\mathcal{M}=\left(
	\begin{array}{cccccc}
		-\kappa_{m} & \Delta_{m} & 0 & \Gamma_{1} & 0 & \Gamma_{2} \\
		-\Delta_{m} & -\kappa_{m} & -\Gamma_{1} & 0 & -\Gamma_{2} & 0 \\
		0 & \Gamma_{1} & -\kappa_{1} & \Delta _{1} & 0 & 0 \\
		-\Gamma_{1} & 0 & -\Delta _{1} & -\kappa_{1} & 0 & 0 \\
		0 & \Gamma_{2} & 0 & 0 & -\kappa_{2} & \Delta _{2} \\
		-\Gamma_{2} & 0 & 0 & 0 & -\Delta _{2} & -\kappa_{2}%
	\end{array}%
	\right). \label{DfM}
\end{equation}%
The Gaussian nature of the system is preserved thanks to the linear dynamics and the Gaussian input noises. Therefore, the steady state of the quantum fluctuations of the system defines the continuous-variable (CV) three-mode Gaussian state.
A $6\times 6$ matrix (CM) V, also known as a covariance matrix, perfectly describes such a state. Its matrix components are as follows:
\begin{equation}
	V_{ij}(t)=\frac{1}{2}\left\langle \mathcal{F}_{i}(t)\mathcal{F}_{j}(t^{\prime })+\mathcal{F}_{j}(t^{\prime})\mathcal{F}_{i}(t)\right\rangle,\quad i,j=1,2,\cdots,6 \label{CoM}
\end{equation}
where the stability conditions of the system derived by using the Routh-Hurwitz criterion are fulfilled (see Appendix A) \cite{72}, the elements of $V$ can directly be achieved by numerically solving the following Lyapunov equation \cite{73}:
\begin{equation}
	\mathcal{M}V+V\mathcal{M}^{T}=-\mathcal{D},  \label{lpEq}
\end{equation}%
where $\mathcal{D}$ is the diffusion matrix that describes the noise correlations. Its elements are defined as:
\begin{equation}
	\mathcal{D}_{ij}(t)\delta(t-t^{\prime })=\frac{1}{2}\left\langle \mathcal{N}_{i}(t)\mathcal{N}_{j}(t^{\prime })+\mathcal{N}_{j}(t^{\prime})\mathcal{N}_{i}(t)\right\rangle,\quad i,j=1,2,\cdots,6 \label{DiffM}
\end{equation}
By using Eqs. (\ref{corrF1})-(\ref{corrF2}), we obtain the explicit expression of the matrix $\mathcal{D}$:
\begin{equation}
	\mathcal{D}=\left(
	\begin{array}{cccccc}
		\kappa_m (2n_m+1) & 0 & 0 & 0 & 0 & 0 \\
		0 & \kappa_m (2n_m+1) & 0 & 0 & 0 & 0  \\
		0 & 0 & \kappa_1(2 N+1) & 0 & 2 M \sqrt{\kappa_1\kappa_2} & 0  \\
		0 & 0 & 0 & \kappa_1(2 N+1) & 0 & -2 M \sqrt{\kappa_1\kappa_2} \\
		0 & 0 & 2 M \sqrt{\kappa_1\kappa_2} & 0 & \kappa_2(2 N+1) & 0 \\
		0 & 0 & 0 & -2 M \sqrt{\kappa_1\kappa_2} & 0 & \kappa_2(2 N+1)
	\end{array}%
	\right). \label{DfM}
\end{equation}
Since the Lyapunov equation (\ref{lpEq}) is a linear equation , it has a straightforward solution. After acquiring the system's CM, it is simple to extract the reduced CMs of the bipartite subsystems (cavity1-cavity2, cavity1-magnon and cavity2-magnon) by neglecting the unwanted columns and rows in the global CM $V$.

In order to quantify the bipartite entanglement in the dual cavity-magnon system, we employ the logarithmic negativity, which is a quantitative measure that captures the Simon criterion \cite{74}. It is defined by \cite{75}:
\begin{equation}\label{LoN}
	E^{N}_{\alpha-\beta}=\max [0,-\ln 2 \eta ^{-}],
\end{equation}%
where $\alpha$ and $\beta$ are any two modes, $\eta ^{-}=$min eig$|i\Omega_2\widetilde{{V}_{4}}|$ is the minimum symplectic eigenvalue of the partially transposed CM $\widetilde{\mathcal{V}_{4}}=\varrho_{1|2}\mathcal{V}_{4}\varrho_{1|2}$, where $\mathcal{V}_{4}$ is a $4\times4$ matrix representing any bipartite subsystem (cavity1-cavity2, cavity1-magnon and cavity2-magnon) and $\varrho_{1|2}={\rm diag}(1,-1,1,1)$ is the $4\times 4$ matrix that defines the partial transposition at the covariance matrices level. Furthermore, $\Omega_2=\bigoplus^{2}_{j=1}i\sigma_{y}$ being a $4\times 4$ symplectic matrix with $\sigma_{y}$ is the $y$-Pauli matrix.

On the other hand, the tripartite entanglement can be described by the minimal residual contangle. The tripartite entanglement remains invariant of  under all possible permutations of the modes. It is defined as \cite{75,76,77}:
\begin{equation}
	\mathcal{R}_{\tau }^{min}\equiv \min [\mathcal{R}_{\tau}^{m|c1c2},\mathcal{R}_{\tau }^{c2|mc1},\mathcal{R}_{\tau }^{c1|c2m}],  \label{RS}
\end{equation}%
where $\mathcal{R}_{\tau }^{u|vw}$ is the the residual contangle given by:
\begin{equation}
	\mathcal{R}_{\tau }^{u|vw}\equiv C_{u|vw}-C_{u|v}-C_{u|w}\geq 0, \ \ (u,v,w=m,c_1,c_2).
\end{equation}
The condition $C_{u|vw}\geq C_{u|v}+C_{u|w}$ is similar to the Coffman-Kundu-Wootters monogamy inequality achieved for the system of three qubits \cite{77}. $C_{x|y}$ is the contagle, which is a proper entanglement monotone, of subsystems $u$ and $v$ ($u$ can contain just one mode, while $y$ can contain one or two modes). The contangle is related to logarithmic negativity and can be expressed as the square of the logarithmic negativity: $C_{x|y}=E_{N,x|y}^{2}$. Noteworthy, for computing the \textit{%
	one-mode-vs-two-modes} logarithmic negativity $E_{u|vw}$, one must adopt the same definition of Eq. (\ref{LoN}), however, the smallest symplectic eigenvalue $\eta
^{-}$ must be modified as:
\begin{equation}
	\eta_{u|vw} ^{-}=\min eig|i\Omega_3\varrho_{u|vw}{V}\varrho_{u|vw}|,
\end{equation}%
where $V$ is the global tripartite covariance matrix, $\Omega_3=\bigoplus^{3}_{j=1}i\sigma_{y}$ is the $6\times 6$ symplectic matrix and $\varrho_{u|vw}$ are the partial transposition matrices given by:
\begin{eqnarray*}
	\varrho_{1|23} &=&\text{diag}(1,-1,1,1,1,1), \\
	\varrho_{2|31} &=&\text{diag}(1,1,1,-1,1,1), \\
	\varrho_{3|12} &=&\text{diag}(1,1,1,1,1,-1).
\end{eqnarray*}	

Owing to the asymmetric nature, the gaussian steering which have differing behaviour than that for entanglement.  It is widely known that entanglement between two modes is a symmetric feature and the entangled state is symmetrically shared between Alice and Bob without specifying the direction, i.e., if Alice is entangled with Bob, Bob is
essentially entangled with Alice. However, this is not true in case of quantum steering due to its asymmetric nature, i.e., Alice may steers Bob but vice versa
may or may not possible
Based on quantum coherent information,
the quantum steerability in different directions, for any two interacting mode Gaussian state, was first introduced by \cite{78}
\begin{eqnarray}
\zeta_{\alpha | \beta } &=&\max \{0,\mathcal{S}(2\mathcal{V}%
_{f})-\mathcal{S}(2\mathcal{V}_{in})\},  \label{A} \\
\zeta_{\beta | \alpha } &=&\max \{0,\mathcal{S}(2\mathcal{V}%
_{g})-\mathcal{S}(2\mathcal{V}_{in})\},  \label{B}
\end{eqnarray}%
where
\begin{equation}
\mathcal{V}_{in}=\left[
\begin{array}{cc}
\mathcal{V}_{f} & \mathcal{V}_{fg} \\
\mathcal{V}_{fg}^{T} & \mathcal{V}_{g}%
\end{array}%
\right]. \label{C}
\end{equation}%
Furthermore, the positivity of the R\'{e}nyi-2 entropy determines the degree of steering,
is defined as $\mathcal{S}(\nu )=\frac{1}{2}\ln \det (\nu )$. Each entry of the above matrix (Eq. (\ref{C})) is a $2\times 2$ matrix. Furthermore, the diagonal
elements $\mathcal{V}_{f}$ and $\mathcal{V}_{g}$ stands for the reduced
state of two modes $f$ and $g$, respectively, while the off-diagonal entries define the correlation between $f$ and $g$. $\zeta_{\alpha | \beta }$ represents the directional EPR steering from mode $\alpha $ to mode $\beta $, while $\zeta_{\beta | \alpha }$ denotes the EPR steering in swaped direction.
No-way steering is observed if $\zeta_{\alpha | \beta }=0$ and $\zeta_{\beta | \alpha }=0$. In addition, the quantification for one-way steering is to have either $\zeta_{\alpha | \beta }>0$ and $\zeta_{\beta | \alpha }=0$ or $\zeta_{\alpha | \beta }=0 $ and $\zeta_{\beta | \alpha }>0$. Finally, if the system has  two-way steering, then $\zeta_{\alpha | \beta }>0$ and $\zeta_{\beta | \alpha }>0$. In addition, asymmetric steerability between any two modes of the Gaussian states can be manipulated by looking at the the steering asymmetry, specified as
\begin{equation}
\zeta_{S}=|\zeta_{\alpha|\beta}-\zeta_{\beta | \alpha}|. \label{ZS}
\end{equation}%
Note that we get one-way or two way gaussian steering as long as steering asymmetry remains positive, i.e., $\zeta_{S}>0$ and no-way gaussian steering if $\zeta_{S}=0$ \cite{79}.
\begin{figure}[b!] \centering\includegraphics[width=1\columnwidth,height=5in]{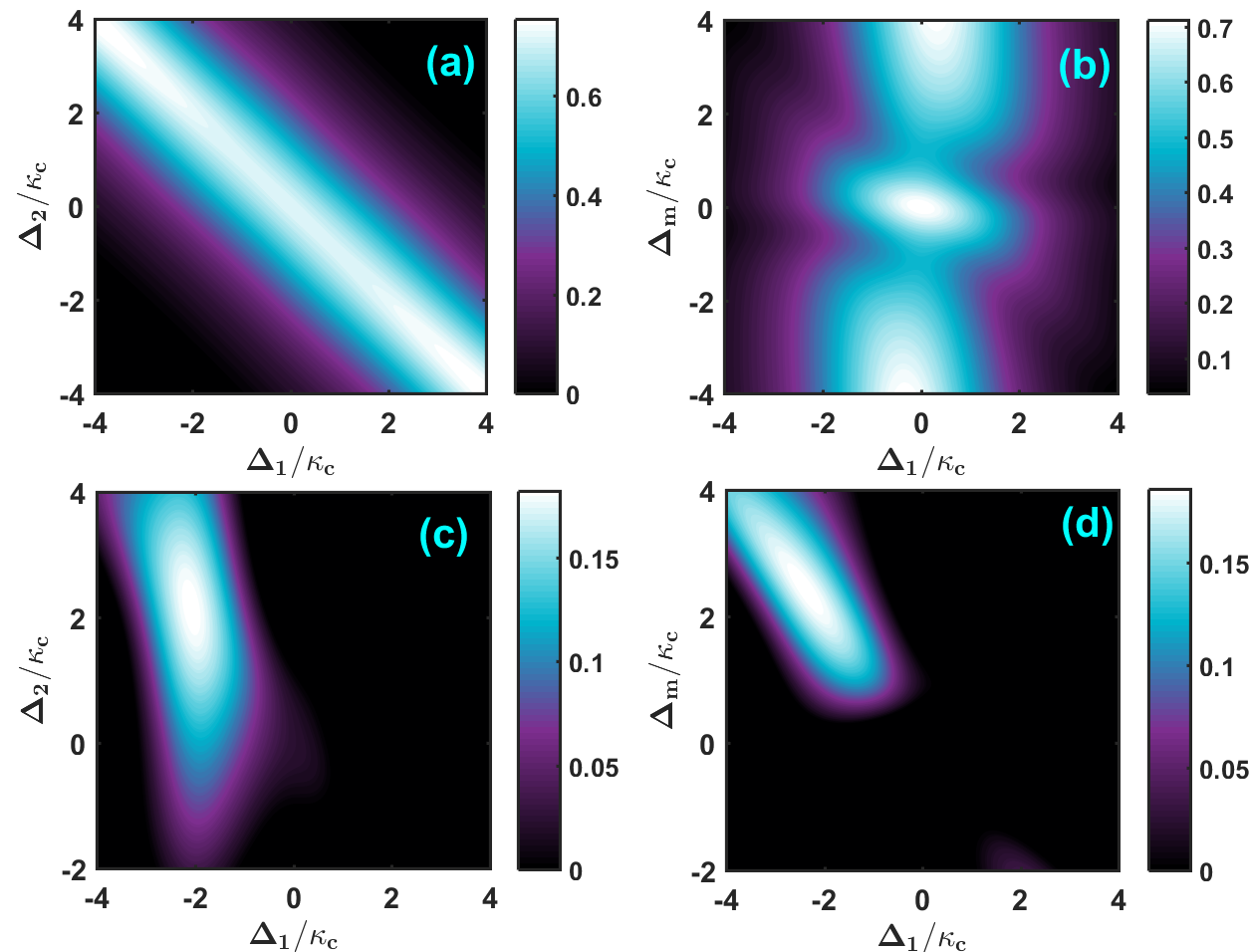}
\caption{ Density plot of the bipartite entanglement: (a) $ E^{N}_{c-c}$ and (c) $ E^{N}_{c-m}$ as a function of $ \Delta_{1}/\kappa_{c}$ and $ \Delta_{2}/\kappa_{c}$ with $ \Delta_{m}=0$ in (a) and $ \Delta_{m}=2 \kappa_{c}$ in (c). Density plot of the bipartite entanglement:  (b) $E^{N}_{c-c} $ and (d) $ E^{N}_{c-m}$ as a function of   $ \Delta_{1}/\kappa_{c}$ and $ \Delta_{m}/\kappa_{c}$ with $ \Delta_{2}=0$ in (b) and $ \Delta_{2}=2\kappa_{c}$ in (d). \small } 		\label{F2}
\end{figure}
\begin{figure}[tbp]
\centering\includegraphics[width=0.95\columnwidth,height=3.2in]{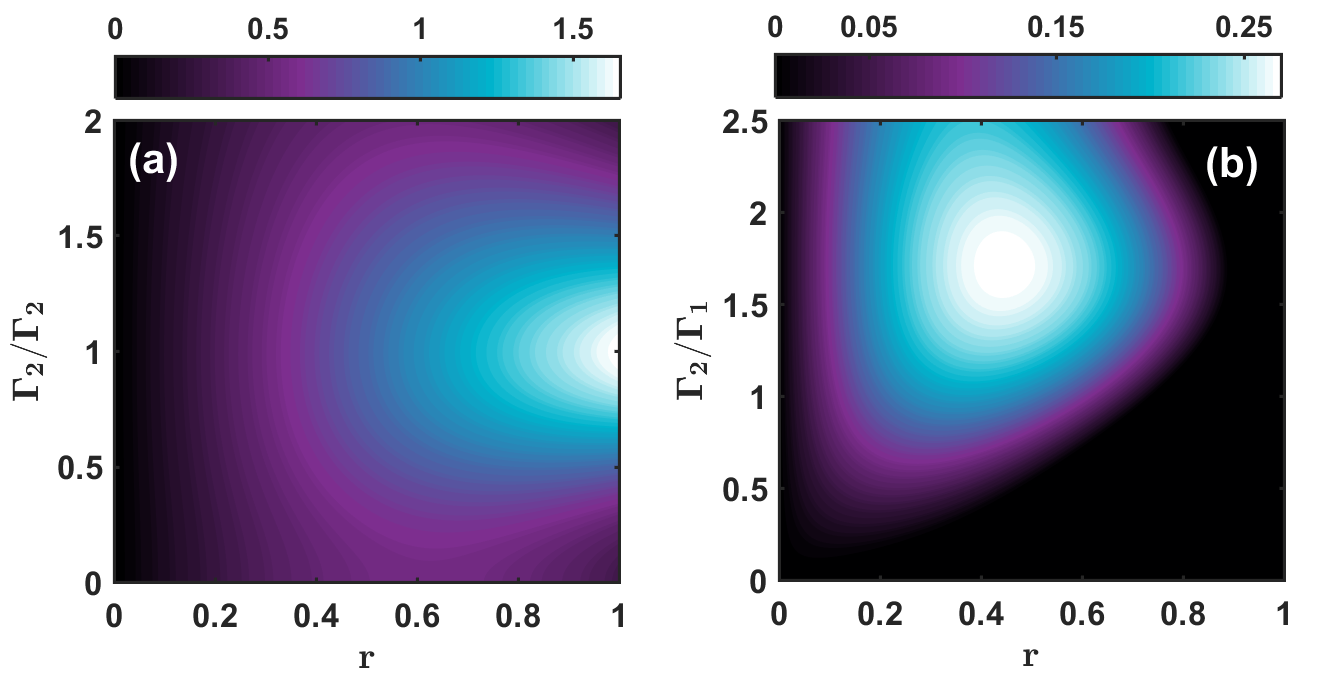}
\caption{ Density plot of the bipartite entanglement: (a) $E^{N}_{c-c}$ and (b) $E^{N}_{c-m}$ as a functions of $r$ and $\Gamma_{2}/\Gamma_{1}$ ($\Gamma_{1}$ is fixed). The parameters are: $ \Delta_{m}=\Delta_{1}=\Delta_{2}=0$ in (a) and $ \Delta_{m}=-\Delta_{1}=\Delta_{2}=2\kappa_{c}$ in (b).
\small }		\label{F3}
	\end{figure}
 \begin{figure}[b!]
\centering\includegraphics[width=0.95\columnwidth,height=3.2in]{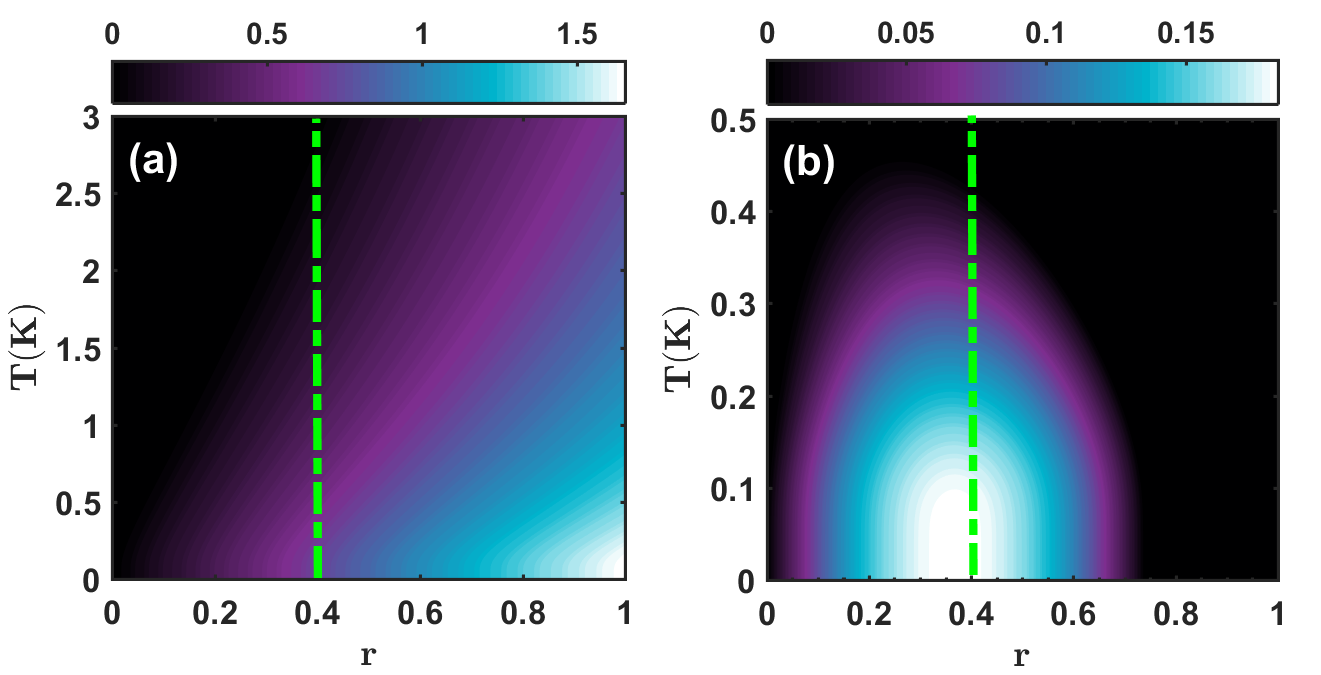}
\caption{ Density plot of the bipartite entanglement: (a) $E^{N}_{c-c}$ and (b) $E^{N}_{c-m}$ as a functions of $r$ and $T$. The parameters are: $ \Delta_{m}=\Delta_{1}=\Delta_{2}=0$ in (a) and $ \Delta_{m}=-\Delta_{1}=\Delta_{2}=2\kappa_{c}$ in (b).
\small }
  	\label{F4}
  \end{figure}
  \begin{figure}[b!]
	\centerline{\includegraphics[width=1\columnwidth,height=5.1in]{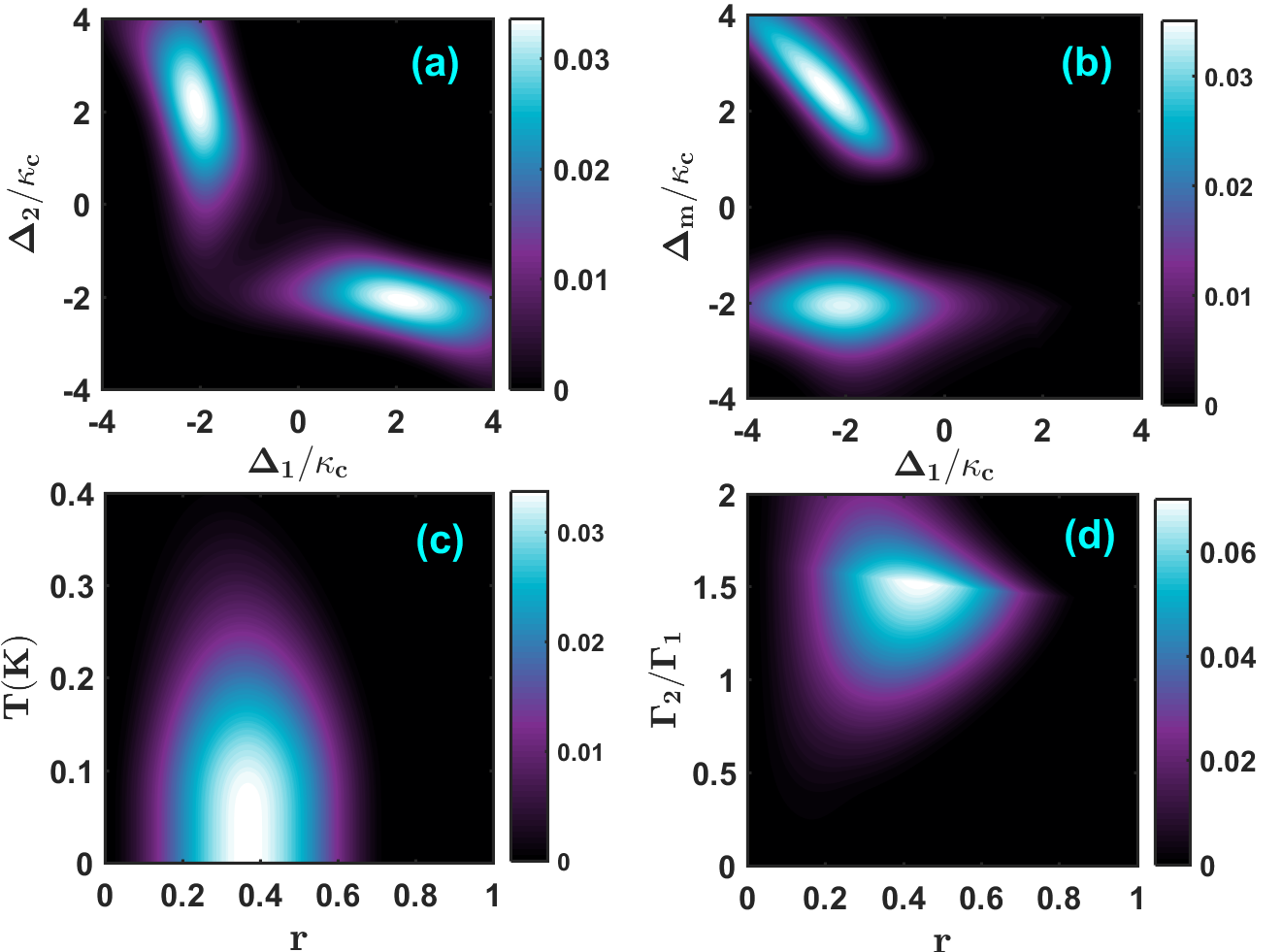}}
	\caption{\footnotesize{Tripartite entanglement  $\mathcal{R}^{min}_{\tau}$  between cavity1-cavity2-magnon modes as a function of (a) $\Delta_{1}/\kappa_c$ and $\Delta_{2}/\kappa_c$, (b) $\Delta_{1}/\kappa_c$ and $\Delta_{m}/\kappa_c$, (c) $ r $ and $ T $, and (d) $ r $ and $ \Gamma_{2}/\Gamma_{1}$. The parameters are:  $\Delta_{m}=2\kappa_c$ in (a),  $\Delta_{2}=2\kappa_c$ in (b), and $\Delta_{m}=-\Delta_{1}=\Delta_{2}=2\kappa_c$ in (c-d).}}
	\label{F5}
\end{figure}
\begin{figure}[tbp]
	\centerline{\includegraphics[width=1\columnwidth,height=2.8in]{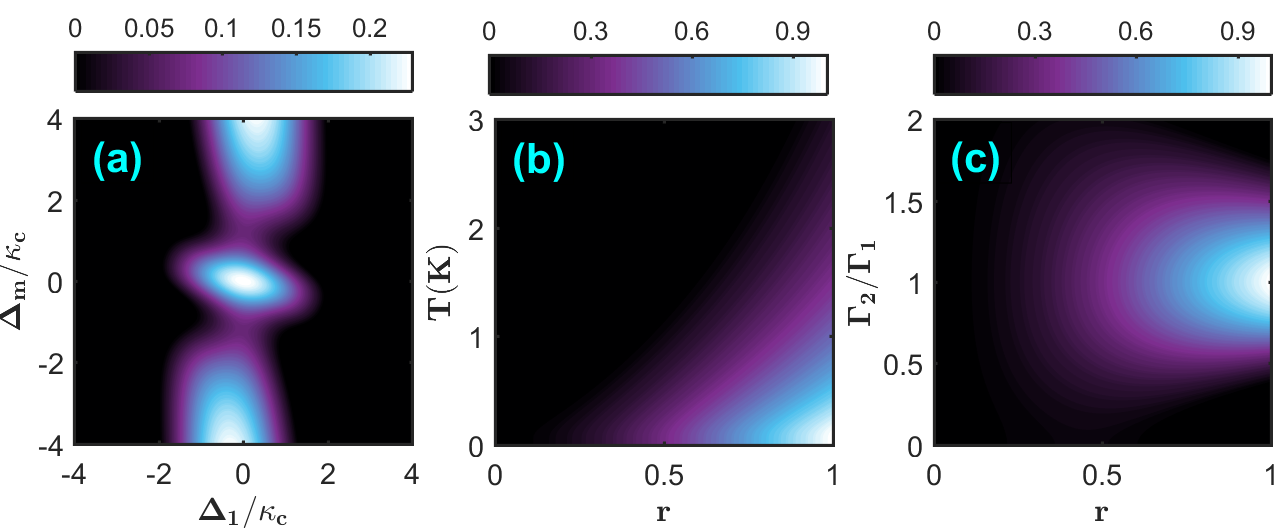}}
	\caption{\footnotesize{Density plots of the gaussian steering $\protect\zeta %
_{c|c}$, versus detunings (a) $\Delta _{1}/\kappa _{c}$ and
$\Delta _{m}/\kappa _{c}$ with $r=0.4$, (b) $r$ and $T$ (c)  $r$ and $ \Gamma_{2}/\Gamma_{1}$. The parameters are same as Fig. 5.}} \label{STE1}
\end{figure}
\begin{figure}[tbp]
	\centerline{\includegraphics[width=1\columnwidth,height=3.1in]{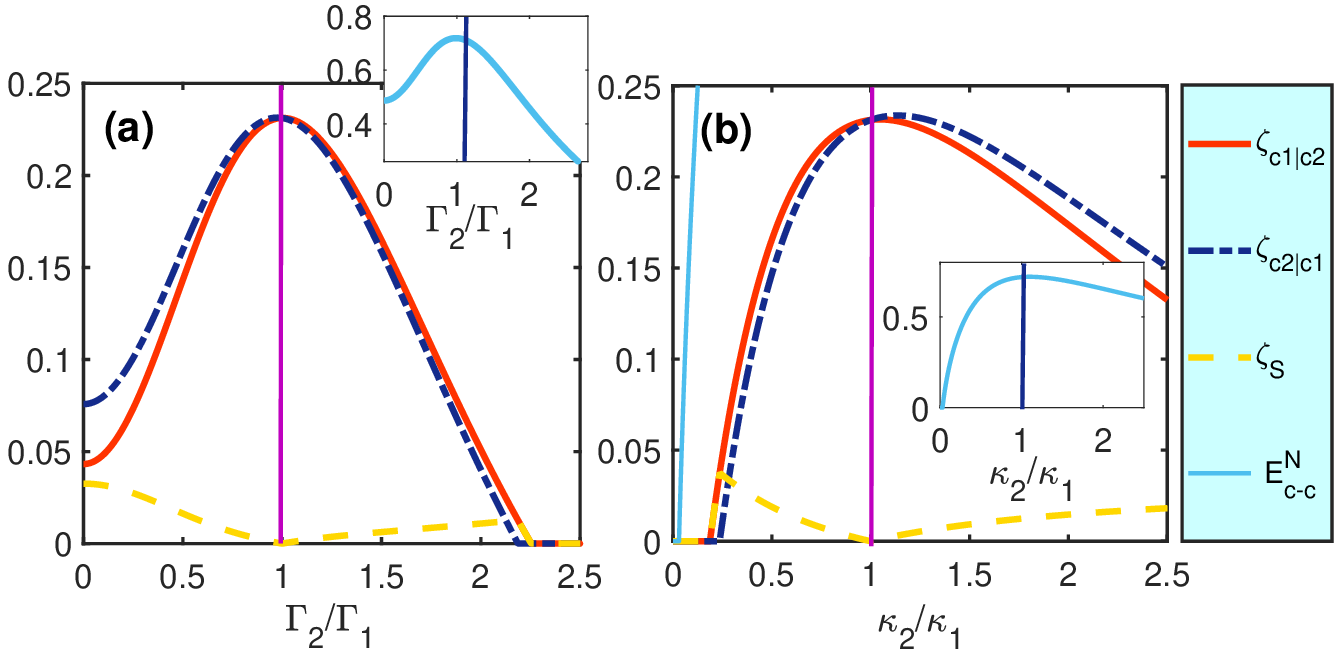}}
	\caption{\footnotesize{Plot of the Gaussian steering $\zeta_{c_2}|\zeta_{c_1}$ (black dashed), $\zeta_{c_1}|\zeta_{c_2}$ (yellow dot-dashed black dashed), $\zeta_{S}$ (magenta solid) and $E_{c-c}^{N}$ (cyan solid in insets) versus detunings (a) $\Gamma _{2}/\Gamma _{1}$ with $\kappa_{1}=\kappa_{2}$ and (b)
$\kappa_{2}/\kappa _{1}$ with $\Gamma _{2}=\Gamma _{1}$. In addition, we take
$\Delta_{1}=\Delta_{2}=\Delta_{m}=0$ and $r=0.4$. The other optimal parameters are as given in text.}} \label{STE2}
\end{figure}

\section{Results and Discussion}
Following, we present our main findings with special emphasis on the bipartite entanglements between two microwave cavities and between cavity and magnon. We consider recent experimentally feasible parameters \cite{39,41}: $\omega_{c}/2\pi= \omega_{m}/2\pi=10 $ GHz,
$\kappa_{c}/2\pi\equiv\kappa_{1}/2\pi=\kappa_{2}/2\pi=5$ MHz , $ \kappa_{m}/2\pi=1$ MHz, $ r=0.4 $,   $ \Gamma\equiv\Gamma_{1}=\Gamma_{2}=4\kappa_{c} $  and $ T=20 $ mK.

Figures 2(a-d) show the bipartite entanglements between the two microwave cavity modes. We observe that the two cavity modes are maximally entangled either the two cavity modes are completely resonant with the magnon mode, i.e., $\Delta_{1}=\Delta_{2}=\Delta_{m}=0$ or at $\Delta_{1}=\Delta_{2}=\Delta_{m}=4\kappa_{c}$ as shown by Fig. 2(a-b). However, we obtained concentrated cavity-magnon entanglement if we increase the cavity and magnon detunings ($\Delta_{m}=2\kappa_{c}$ in  Fig. 2(c) and $\Delta_{2}=2\kappa_{c}$ in  Fig. 2(d)) as shown by Fig. 2(c-d). Note that in this case, the two cavity modes must be in resonance with the two sidebands of the magnon mode, respectively, i.e., $-2\Delta_{1}=2\Delta_{2}=2\Delta_{m}$.

Next, we discuss the effect of the mismatch of the two couplings on the current scheme
Nair, et. al. \cite{35} recently proposed a system to give rise  entanglement between two magnon modes using a single-mode squeezed field. They have demonstrated that the optimal entanglement between two magnon modes can be established  only if one of the collective quadrature of two magnon modes is squeezed while taking equal amount of photon-magnon coupling strengths. However, we use a different scheme to entangle two cavity modes and cavity-magnon modes using the quantum correlation transferred from a squeezed field. Note that Fig. \ref{F3} (a) shows that, with the rise of the squeezing coefficient $r$, considerable entanglement between two cavity modes $E_{c-c}^{N}$ is produced for a broad range of mismatch of the two photon-magnon coupling strengths. Moreover, it can be seen that smaller value of squeezing parameter $r$ is much more permissive to the mismatch. However, the entanglement between cavity and magnon $E_{c-m}^{N}$ is concentrated for a specific range of squeezing parameter and optimal entanglement is achieved around $r=0.45$ as exhibited by Fig. \ref{F3} (b).

It is vital to mention that we have considered   the cavity linewidths is smaller than bandwidths of input squeezed fields. In Ref. \cite{67}, logarithmic negativity $E^N=0.8$ which corresponds to the bandwidth of 12.5 MHz and squeezing r = 0.4 has been generated for a two-mode squeezed field.
Employing the same experimental parameters, we have achieve cavity-cavity entanglement $E_{c-c}^{N}=0.7$ and cavity-magnon entanglement $E_{c-m}^{N}=0.18$ at $T=20$mK as shown by Fig. \ref{F4} (a-b). Furthermore, it is vital to note that $E_{c-c}^{N}$ ($E_{c-m}^{N}$) survives upto larger temperature than our previous findings \cite{45}, i.e., upto $T\sim2.8$K ($T\sim 0.4$K) which turns out to be much simpler and better than the systems based on nonlinearities \cite{49}.
Figure \ref{F4} (a-b) demonstrates how increasing the squeezing parameter $r$ causes an increase in the average photon number in the squeezed field, which initially improves cavity-magnon entanglement before being linearly shifted to enhance the entanglement between two MW cavities. However, this enhancement is finite to specific region. Furthermore, the bipartitions $E_{c-c}^{N}$ and $E_{c-m}^{N}$ decreases with the enhancement of temperature.

The creation of photon-magnon-photon genuine tripartite entanglement is a eminent feature in the cavity-magnon system. To measure genuine tripartite
entanglement, we have considered the minimum residual cotangle criterion. In Fig. \ref{F5}, we demonstrate genuine tripartite entanglement (photon-magnon-photon) by the nonzero (positive) minimum residual contangle $\mathcal{R}_{\tau }^{min}$. It clear from Fig. \ref{F5} (a-b) that the genuine tripartite photon-magnon-photon entanglement is maximally found when the two cavity modes are almost resonant with the magnon mode sidebands, respectively, i.e., at $-\Delta_{1}=\Delta_{2}=\Delta_{m}$ or $\Delta_{1}=-\Delta_{2}=\Delta_{m}$.
It can be seen in Fig. 5 (c-d) that the genuine tripartite entanglement has been significantly altered by the squeezing parameter $r$. In addition, Fig. \ref{F5} (c) shows the genuine photon-magnon-photon tripartite entanglement versus temperature and squeezing parameter, clearly show that it exist for a specific range of squeezing parameter i.e., $0.1 \leq r \leq 0.6$. In addition, \ref{F5} (c) clearly showing that tripartite entanglement cease to exist around $380$mK for $r=0.4$, which is even larger than the previous findings of generating tripartite entanglement without JPA \cite{65}. Hence, JPA in present scheme plays a key role in enhancing photon-magnon-photon tripartite entanglement.

In the conversation that follows is to discuss quantum steering in detail. It is well known that steerable states between any two modes
are always entangled, but, the converse is not generally legitimate, that is, there is no restriction on entangled states to be steerable. This feature endorse the idea that there must be stronger correlations to realize the Gaussian steering.
In Fig. 6, we display an important resulting showing two-way steering between indirectly coupled cavity modes, however, we have not observe the steering between magnon and cavity modes. It is important to mention here that as we have taken the same parameters for the two cavities, therefore, the two identical cavity modes equally steer each other, i.e., $\zeta_{c_{_1} | c_{_2} }=\zeta_{c_{_2} | c_{_1} }=\zeta_{c | c }$. It is shown that both cavity modes maximally steer each other either at the resonance point  $\Delta _{1}=\Delta _{2}=\Delta _{m}=0$ or far from the resonance point $\Delta _{1}=\Delta _{2}=0$, $\Delta _{m}=\pm 4\kappa _{c}$ as shown in Fig. 6 (a). The two-way steering implies that Alice and Bob, as the two parties, can assure one another that they are sharing an entangled state.
It is also shown in Fig. 6 (b) that not only the steerability but also the survival of steerability increase with the enhancement of squeezing parameter $r$, which suggests that
Gaussian steering is noticeably sensitive to thermal noise. In addition, we obtained optimal directional steerability when the two optomagnonical couplings become equal for the large value of squeezing parameter $r$ as shown in Fig. 6 (c).

In Fig. 7 (a-b), we plot the directional steering  $\zeta_{c_1}|\zeta_{c_2}$ (red solid), $\zeta_{c_2}|\zeta_{c_1}$ (blue dot-dashed), steering symmetry $\zeta_{S}$ (yellow dashed) and entanglement $E_{c-c}^{N}$ (cyan solid) to discuss the effect of unequal magnomechanical coplings and decay rates on steerability and entanglement between the two cavity modes.
Since steerable states are always entangled, therefore, strong quantum correlations are
obligatory for realizing the quantum steering than that for the entanglement.
It can be seen from Fig. \ref{STE2}(a) that $\zeta_{c_2|c_1}>\zeta_{c_1|c_2}$ for $\Gamma_{2}<\Gamma_{1}$ and $\zeta_{c_2|c_1}<\zeta_{c_1|c_2}$ for $\Gamma_{2}>\Gamma_{1}$, however, $\zeta_{c_2|c_1}=\zeta_{c_1|c_2}$ for $\Gamma_{2}=\Gamma_{1}$. In addition, Fig. \ref{STE2}(b) shows that $\zeta_{c_2|c_1}>\zeta_{c_1|c_2}$ for $\kappa_{2}>\kappa_{1}$ and $\zeta_{c_2|c_1}<\zeta_{c1|c2}$ for $\kappa_{2}<\kappa_{1}$, however, $\zeta_{c_2|c_1}=\zeta_{c_1|c_2}$ for $\kappa_{2}=\kappa_{1}$. It is important to mention here that a crucial and  significant tool for safe quantum teleportation of continuous-variable systems is two-way Gaussian steering.

\section{Conclusions}\label{sec5}
We have presented a protocol to investigate the enhancement of entanglement of mainly two kinds bipartitions via JPA, namely cavity-cavity and cavity-magnon, in a dual cavity magnon system. A magnon mode is simultaneously coupled with the two cavity modes via magnetic dipole interaction. In this paper, we have theoretically studied that the steady-state entanglements in the current system can be achieved within the experimentally feasible parameters.
The entanglements $E^{N}_{c-c}$ and $E^{N}_{m-c}$ were found to be robust against ambient temperature and survive up to $T\sim2.8$K and $T\sim 0.4$K, respectively. In addition, we have also examined the effectiveness of two entanglements under the situation of mismatch of optomagnonical couplings of magnon with both cavities and found that cavity-cavity entanglement
transfer efficiency can be attained by substantial squeezing parameter and optomagnonical coupling strengths.
Furthermore, the correlation degree of tripartite entanglement not only found to be controllable via adjusting the system parameters such as cavity and magnon detunings, the cavity-magnon optomagnonical coupling strengths, and squeezing parameter, but also robust against the temperature. Our proposal for generating entanglement between two indirectly cavity modes does not demand any nonlinearities and, therefore, goes against the conventional sense of improving entanglement. Finally, the quantum steering between two indirectly coupled mode is achieved at the resonance point with equal magnomechanical couplings and decay rates.
Our scheme broadens the horizons to manipulate the quantum correlation in hybrid quantum systems.
\begin{figure}[b!]
\centerline{\includegraphics[width=0.95\columnwidth,height=2.8in]{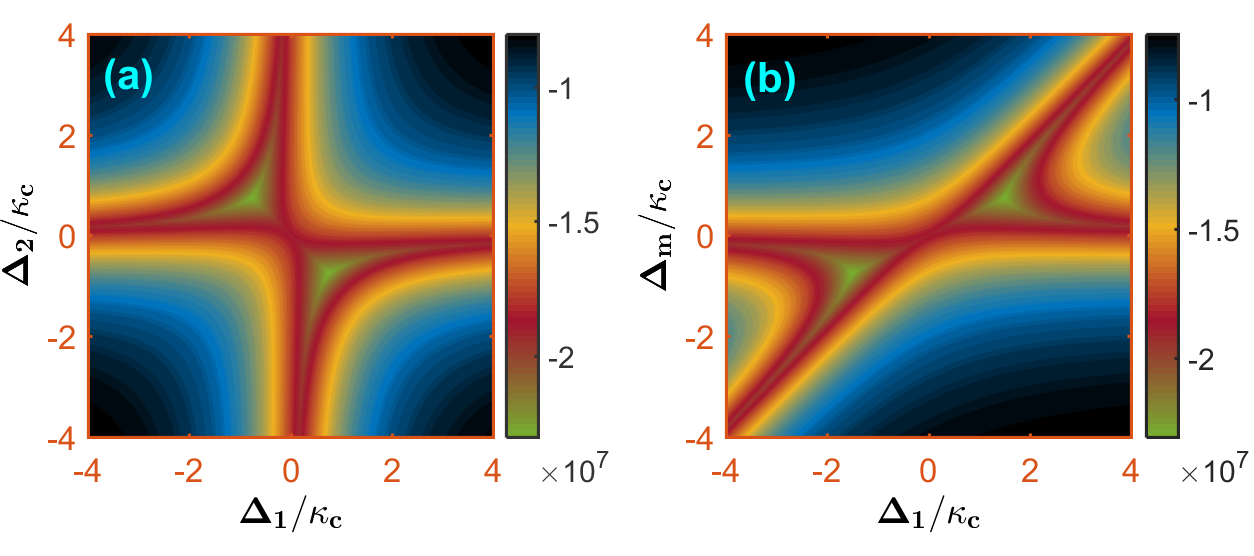}}
\caption{\footnotesize{The maximum of the real parts of the eigenvalues obtained from the reduced drift matrix for (a) cavity-cavity (b) cavity-magnon vs normalized cavity detunings.}}
	\label{EV}
\end{figure}

\section*{Appendix A. Stability of the system}
Discussing the stability of the existing system is essential since stability in the cavity-magnon system is challenging to accomplish. In general, the system only demonstrates stability when all of the negative real components eigenvalues of the drift matrix $\mathcal{M}$ are obtained. However, a system only exhibits instability if any real part of the eigenvalue has a positive sign. The Routh-Hurwitz criterion, which completely ensures the stability of any system, is therefore adopted in this context \cite{80}. In Fig. \ref{EV}, we show the graphical depiction of the eigenvalues (maximum of the real parts) $\lambda_{\mathcal{M}}$ ($|\mathcal{M}-\lambda_{\mathcal{M}}\mathbb{1}|=0$) of the drift matrix $\mathcal{M}$ vs. the normalized detunings.
By employing the chosen parameters for the current system, Fig. \ref{EV} unequivocally shows that the system is stable because the maximum of the real parts of the eigenvalues remains negative.
Furthermore, It can be seen in Fig. \ref{EV} (a-b) that the system is more stable around the resonance points than surrounding region.
Conclusively, the experimental parameters utilized throughout the simulations fulfills the stability conditions required by the Routh-Hurwitz criterion and therefore, the working regime is the regime of stability.
\section*{Data availability}
All data used during this study are available within the article.

\end{document}